\documentclass[a4paper]{jpconf}
\usepackage{graphicx}
\usepackage{amsmath, amssymb, amsthm, amscd}
\usepackage{bbm}
\usepackage{mathrsfs,mathtools}
\usepackage{tikz}
\usepackage{float}
\usepackage[utf8]{inputenc}
\usepackage[english]{babel}
\newcommand\sS{\mathbb{S}}
\newcommand\hh{\mathbb{H}}
\newcommand{\dd}{\mathrm{d}}
\newcommand{\h}{\mathrm{h}}
\usepackage{mathtools}
\usepackage{amsfonts,amssymb,amsthm}

\usepackage{xcolor}

\newcommand\bbbone{{\mathchoice {\rm 1\mskip-4mu l} {\rm 1\mskip-4mu l}
{\rm 1\mskip-4.5mu l} {\rm 1\mskip-5mu l}}}
\newcommand{\one}{\bbbone}

\newcommand\rr{\mathbb{R}}
\newcommand{\ii}{\mathrm{i}}

\renewcommand{\e}{\mathrm{e}}

\renewcommand{\Re}{\operatorname{Re}}

\newcommand{\gen}{\mathrm{gen}}

\DeclarePairedDelimiterX{\cinner}[2]{(}{)}{#1\,\delimsize\vert\,\mathopen{}#2}
\DeclarePairedDelimiterX{\rinner}[2]{\langle}{\rangle}{#1\,\delimsize\vert\,\mathopen{}#2}

\newcommand{\s}{{\mathrm s}}

\renewcommand{\mid}{\mkern5mu|\mkern5mu}

\newcommand{\aotimes}{{\mathop{\otimes}\limits^{
  \vbox to .15ex {\kern-2\ex@\hbox{\tiny alg}\vss}}}}

\renewcommand{\bar}{\overline}

\newcommand{\cW}{{\mathcal W}}

\newcommand{\cC}{{\mathcal C}}
\newcommand{\cB}{{\mathcal B}}
\newcommand{\cG}{{\mathcal G}}

\newcommand{\bes}{\begin{subequations}}
\newcommand{\ees}{\end{subequations}}

\begin{document} 

\title{Generalized integrals and point interactions}
 
 \author{Jan Derezi\'{n}ski$^{1}$,
Christian Ga\ss{}$^1$ and B{\l}a{\.z}ej Ruba$^2$}

\address{$^{1}$ Department of Mathematical Methods in Physics, Faculty of Physics, 
University of Warsaw, Pasteura 5, 02-093 Warszawa, Poland}
\address{$^2$ Department of Mathematics, University of Copenhagen, Universitetsparken 5, DK-2100 Copenhagen Ø, Denmark}

\ead{jan.derezinski@fuw.edu.pl}

\begin{abstract}
First we recall a method of computing scalar products of
eigenfunctions of a Sturm-Liouville operator. This method is then applied to
Macdonald and Gegenbauer functions, which are eigenfunctions of the
Bessel, resp. Gegenbauer operators. 
 The computed scalar products are well defined only
for a limited range of parameters. To extend the obtained formulas
to a much larger range of parameters, we introduce the concept of a
generalized integral. 
The (standard as well as generalized) integrals of Macdonald and
Gegenbauer functions
have important applications to operator theory. Macdonald functions
can be used to express  the
integral kernels
of the resolvent (Green functions) of the Laplacian on the Euclidean
space in any dimension. Similarly, Gegenbauer
functions appear in Green functions of the Laplacian on the sphere
and the hyperbolic space. In dimensions 1,2,3 one can perturb these
Laplacians with a point potential, obtaining a well defined
self-adjoint operator. Standard integrals of Macdonald and Gegenbauer
functions appear in the formulas for the corresponding  Green functions.
In  higher dimensions the Laplacian perturbed by point potentials  does
not exist. However, the corresponding Green function can be generalized to any dimension by using generalized integrals. 
\end{abstract}

The following notes are a short version of our papers
\cite{DeGaRu1,DeGaRu2}.

\section{Bilinear integrals of eigenfunctions of Sturm-Liouville operators}
Consider a {\em Sturm-Liouville operator}
\begin{align*} 
\cC:=-\rho(r)^{-1}\Big(\partial_rp(r)\partial_r+q(r)\Big)
\end{align*} 
acting on functions on an interval $]a,b[$. $\cC$ is formally
{\em symmetric} for the bilinear scalar product with the {\em density} $\rho$:
\begin{align*}\langle f|g\rangle:=\int_a^b f(r)g(r)\rho(r)\dd r.
\end{align*}
Let us describe a method to compute the scalar product of two
{\em eigenfunctions} of $\cC$.

First, consider  eigenfunctions $f_i$ corresponding to two {\em
  distinct
eigenvalues} $ E _i$, $i=1,2$.
That is, suppose that $ \cC f_i= E _i f_i.$
Then the following is true:
\begin{align}
\int_a^b f_1(r)f_2(r)\rho(r)\dd
                 r&=\frac{\cW(b)-\cW(a)}{E_1-E_2}\label{green}\\
\text{where } \cW(r):=
     f_1(r) p(r)f_2'(r) &
                                    -f_1'(r)p(r)f_2(r) \text{ is the
                       \em  Wronskian}.\notag\end{align}
        \eqref{green} is sometimes called
{\em                          Green's identity}, or the {\em integrated  Lagrange identity}. 
Note that if $a,b$ are {\em singular points} of the
 corresponding differential equation, then the right hand side  of
 \eqref{green} can  often be easily evaluated.


Using  an appropriate limiting procedure we can also often
evaluate \eqref{green}  for $f=f_1=f_2$:
\begin{align}\label{square}
\langle f|f\rangle=\int_a^b f(r)^2\rho(r)\dd r.\end{align}

\section{Bilinear integrals of Macdonald and Gegenbauer functions}

The following two families of Sturm-Liouville operators 
are especially important for applications \cite{WW,Olver,NIST}: 
\begin{align*}
\text {the \em Bessel operator:}\qquad\cB_\alpha & :=-\frac1r\partial_rr\partial_r+\frac{\alpha^2}{r^2}, 
\\\text {and the \em Gegenbauer operator}: \quad
\cG_{\alpha} & := - (1-w^2)^{-\alpha}\partial_w(1-w^2)^{\alpha+1}\partial_w. 
\end{align*}

The   {\em modified Bessel equation} is the eigenequation
of $\cB_\alpha$
with eigenvalue $-1$. The  {\em (standard) Bessel
equation} is its eigenequation for  eigenvalue $1$.
The separation
of variables in the Laplacian
on the {\em Euclidean space}  $\rr^d$ leads to the Bessel operator
on $[0,\infty[$ with  density $\rho=2r$ and $\alpha=\frac{d-2}{2}$.

The {\em Gegenbauer equation} is the eigenequation of
$\cG_\alpha$ with eigenvalue 
$\lambda ^2-\big(\alpha +\frac{1}{2}\big)^2$. The Gegenbauer operator 
on $[-1,1]$ with density  $\rho(w)=(1-w^2)^\frac{d-2}{2}$ 
 arises when we separate variables 
 in the Laplacian on the {\em sphere} $\sS^d$.
 We obtain the Gegenbauer operator 
on $[1,\infty[$ with density  $\rho(w)=(w^2-1)^\frac{d-2}{2}$ 
  when we separate variables 
 in the Laplacian on the {\em hyperbolic space} $\hh^d$.


Scaled {\em Macdonald functions}
$K_\alpha(br)$ are exponentially decaying eigenfunctions of the Bessel operator with
eigenvalue $-b^2$.
Applying the above method  for
$a>0$, $b>0$ we  find
 \begin{eqnarray}
\int_0^\infty  K_\alpha (ar) K_\alpha (br)2r\dd r
&=& \frac{\pi\big((a/b)^{\alpha }-(b/a)^{\alpha }\big)}{\sin(\pi\alpha)(a^2-b^2)}, \quad\begin{array}{c} |\Re(\alpha)|<1,\\\alpha \neq0;\end{array}\label{mac1}\\
 \int_0^\infty  K_0(ar) K_0(br)2r\dd r
&=&\frac{2 \ln \frac{a}{b}}{a^2-b^2},\label{mac2}\\
\label{int3} \int_0^\infty  K_\alpha (br)^22r\dd r&=&\frac{\pi \alpha }
{b^2\sin (\pi \alpha ) }, \qquad \begin{array}{c} |\Re(\alpha)|<1,\\\alpha \neq0;\end{array}\label{mac3}\\
\label{int4} \int_0^\infty  K_0(br)^22r\dd
r&=&\frac{1}{b^2}.\label{mac4}
\end{eqnarray}

The first  formula follows directly by Green's identity.
The next three identities are obtained by                            
applying the {\em de l'H\^opital rule}  to $\alpha=0$ and $a=b$.

The identities  \eqref{mac1}--\eqref{mac4} can be found in standard collections of 
integrals, such as \cite{GraRy}.

The {\em Gegenbauer equation} is  the special case of the hypergeometric
equation
with the symmetry $w\to-w$ and the singular points at $-1,1,\infty$:
\begin{align*}
\Bigg((1-w^2)\partial_w^2-2(1+\alpha )w\partial_w
+\lambda ^2-\Big(\alpha +\frac{1}{2}\Big)^2\Bigg)f(w)=0.
\end{align*}
 In the present context, the Gegenbauer equation is arguably more convenient 
than the equivalent, but more frequently encountered {\em associated Legendre 
equation}.

We will use two kinds of {\em Gegenbauer functions}: one
 is  characterized by its asymptotics $\sim \frac{1}{\Gamma(\alpha+1)}$ at $1$:
\begin{align*}
{\bf    S}_{\alpha ,\pm \lambda }(w)&
    := \sum_{j=0}^\infty
    \frac{\big(\frac12+\alpha+\lambda\big)_j\big(\frac12+\alpha-\lambda)\big)_j}
    {\Gamma(\alpha+1+j)j!}\Big(\frac{1-w}{2}\Big)^j,
\end{align*}
where $(z)_n$ is the \emph{Pochhammer symbol.} 
The other has the asymptotics $\sim \frac{1}{w^{\frac12+\alpha+\lambda}\Gamma(\lambda+1)}$ at $\infty$:
\begin{align*}
 \mathbf{Z}_{\alpha ,\lambda }(w):
 &=
 \frac{1 }{ ( w \pm 1)^{\frac12+\alpha +\lambda }}
\sum_{j=0}^\infty\frac{\big(
\frac12+\lambda\big)_j\big(\frac12+\lambda+\alpha\big)_j}{\Gamma(\lambda+1) (1+2\lambda)_jj!}\Big(\frac{2}{1\pm w}\Big)^j.
\end{align*}

We note the identities

\begin{align*}
  {\bf S}_{\alpha,\lambda}(w)={\bf S}_{\alpha,-\lambda}(w),&\quad
     {\bf Z}_{\alpha,\lambda}(w)=\frac{ {\bf Z}_{-\alpha ,\lambda }(w)
     }{(w^2-1)_\bullet^\alpha},
  \end{align*} 
as well as the slightly more subtle {\em Whipple identity}:
 
  \begin{align*}
 {\bf Z}_{\alpha ,\lambda }(w)&:=  (w^2-1)_\bullet^{-\frac14-\frac\alpha2 
                                -\frac\lambda2}{\bf S}_{\lambda 
                                ,\alpha 
                                }\left(\frac{w}{(w^2-1)_\bullet^{\frac12}}\right),\\
 {\bf S}_{\alpha ,\lambda }(w)&:=  (w^2-1)_\bullet^{-\frac14-\frac\alpha2 
                                -\frac\lambda2}{\bf Z}_{\lambda 
                                ,\alpha 
                                }\left(\frac{w}{(w^2-1)_\bullet^{\frac12}}\right), \qquad \Re(w) >0,
  \end{align*}
  where $(w^2-1)_\bullet^\alpha := (w-1)^\alpha (w+1)^\alpha$ and we
  use the principal branch of the power function.

  Here are the basic bilinear integrals of Gegenbauer functions.
We assume $|\Re(\alpha)|<1$, $\alpha\neq0$ and $\Re(\lambda)>0$:
  \begin{align}\label{geg1}
&\int_{-2}^2{\bf S}_{\alpha,\ii\beta_1}(w){\bf
  S}_{\alpha,\ii\beta_2}(w)(1-w^2)^\alpha\dd 2w
\\=&   \frac{2^{2 \alpha +2}}{(\beta_1^2-\beta_2^2)\sin\pi\alpha} \Big(\frac{\cosh(\pi\beta_1)
}{\Gamma(\frac12+\alpha-\ii\beta_2) 
  \Gamma(\frac12+\alpha+\ii\beta_2)}
  -(\beta_1\leftrightarrow\beta_2)\Big)\notag
\end{align}
  \begin{align}\label{geg2}
    &
\int_2^\infty{\bf Z}_{\alpha,\lambda_1}(w){\bf
  Z}_{\alpha,\lambda_2}(w)(w^2-1)^\alpha\dd 2w\\=&
   \frac{2^{\lambda_1+\lambda_2+1}
                  }{(\lambda_1^2-\lambda_2^2)\sin\pi\alpha}\Big(
                 \frac{1}{\Gamma(\frac12-\alpha+\lambda_1) 
          \Gamma(\frac12+\alpha+\lambda_2)}
          -(\lambda_1\leftrightarrow\lambda_2)\Big)
               \notag                                    . \end{align}

Applying the de l'H\^opital rule we extend these identities to $\alpha=0$,
$\lambda_1=\lambda_2$, and $\beta_1=\beta_2$, see \cite{DeGaRu1}.  
In contrast to the  well-known integrals over Macdonald functions, the integrals 
\eqref{geg1} and \eqref{geg2} seem to be new.

\section{Generalized integral}
 
The integrals \eqref{mac1}--\eqref{mac4} and
\eqref{geg1}--\eqref{geg2} are divergent for $|\Re(\alpha)|\geq1$. 
 Using the  {\em generalized integral}, 
\eqref{mac1}--\eqref{mac4} and \eqref{geg2} can be extended to 
all $\alpha\in\mathbb{C}$, and \eqref{geg1} can be extended to 
$\Re(\alpha)>-1$. The concept of the generalized integral can be 
traced back to  Hadamard \cite{Hadamard23,Hadamard32} and  Riesz 
\cite{Riesz}; see \cite{Lesch97,Paycha} for modern expositions.

Note that several variations of the generalized integral are 
possible. In our note we restrict ourselves to functions which 
are non-integrable near a finite point and have a finite number 
of homogeneous singularities at this point. In some 
of the above references one can also find other variations of the
generalized integral applicable to functions which are non-integrable 
near infinity and/or exhibit \emph{almost homogeneous} singularities, i.e., singularities that are homogeneous up to powers of 
logarithms. 

 Without loss of generality, we can put the singular point to 0. We then say 
 that a function 
$f$ on $]0,\infty[$ is {\em integrable in the generalized sense} if it is 
integrable on $]1,\infty[$ and if there exists a finite set $\Omega\subset\mathbb{C}$ 
and  $f_k\in\mathbb{C}$, $k\in\Omega$, such that
$ f-\sum\limits_{k\in\Omega} f_kr^k$
is integrable on $]0,1[$.  We define the {\em generalized integral} as
\begin{align*}
\gen\int_0^\infty f(r)\dd r
:= &\sum_{k\in\Omega\backslash\{-1\}}\frac{f_k}{k+1}+
\int_0^{1}\Big(f(r)-\sum_{k\in\Omega}f_k r^k\Big)\dd r+\int_{1}^\infty
f(r)\dd r.\end{align*}
For $f\in L^1[0,\infty[$ the generalized and standard integrals
coincide:
\begin{align*}
\gen\int_0^\infty f(r)\dd r=\int_0^\infty f(r)\dd r.\end{align*}

If $f_{-1}\neq0$, the generalized integral has a {\em scaling
  anomaly} but it is always invariant with respect to a power transformation:
\begin{align*}
\gen\int_0^\infty f(r)\dd r
&=\gen\int_{0}^\infty f(\alpha u)\,  \alpha \dd u +f_{-1}\ln(\alpha), 
\quad
\\\qquad
 \gen\int_0^\infty f(r)\dd r
&=\gen\int_{0}^\infty f( u^\alpha)\, \alpha u^{\alpha-1}\dd u.
\end{align*}

Under more general coordinate transformations $g(u)$, which are smooth and  
preserve the leading scaling behavior of the integrand (i.e., $g(0)=0$ and 
$g'(0)\neq0$), the generalized integral transforms as 
\begin{align}
    & \gen \int_0^\infty  f(g(u))g'(u) \dd u - \gen \int_0^\infty  f(r) \dd r  \label{eq:change_of_var} \\ \notag
    = &- f_{-1} \ln g'(0) + \sum_{l\in(\mathbb{N}+1)\cap\Omega}
    \frac{f_{-l}}{(l-1)(l-1)!} \left. \frac{\dd^{l-1}}{\dd u^{l-1}}  \left( \frac{u}{g(u)} \right)^{l-1} \right|_{u=0}.
\end{align}
That is to say, the generalized integral transforms non-trivially under such 
transformations if there is an $n\in\mathbb{N}$ such that $f_{-n}\neq0$.
For this reason, we call the generalized integral {\em anomalous} if $f_{-n}\neq0$
for some $n\in\mathbb{N}$.

Non-anomalous generalized integrals have much better properties than 
anomalous ones. They are often easy to  compute: one just applies analytic
continuation.

A systematic analysis of properties of the generalized integral and a proof 
of \eqref{eq:change_of_var} can be found in \cite{DeGaRu1}. To our knowledge, 
the formula \eqref{eq:change_of_var} is new.

\section{Bilinear generalized integrals of Macdonald and Gegenbauer functions}

Let $a,b>0$. For $\alpha\not\in\mathbb{Z}$ generalized integrals
of Macdonald functions are analytic
continuations
of standard integrals:
\begin{align*}
\gen\int_0^\infty  K_\alpha (ar) K_\alpha (br)2r\dd r
&=\frac{\pi}{\sin(\pi \alpha )}
\frac{\big(\frac{a}{b}\big)^{\alpha }-\big(\frac{b}{a}\big)^{\alpha }}{a^2-b^2}, \\
 \gen\int_0^\infty  K_\alpha (br)^22r\dd r&=\frac{\pi \alpha }
{b^2\sin (\pi \alpha ) }.
\end{align*}
 They have poles at $\alpha\in\mathbb{Z}$.

For $\alpha\in\mathbb{Z}$ the generalized integrals are anomalous and
more complicated to compute. In particular, they do not coincide with
the {\em finite parts} of the above expressions:
\begin{align*}
 \gen\int_0^\infty  K_\alpha (ar) K_\alpha (br)&2r\dd r
=(-1)^\alpha 2
\frac{\big(\frac{a}{b}\big)^\alpha\ln\big(\frac{a}{2}\big) -\big(\frac{b}{a}\big)^\alpha\ln\big(\frac{b}{2}\big) }{a^2-b^2}\\
 &\hspace{-18ex} -\frac{(-1)^\alpha }{ab} \sum_{k=0}^{|\alpha |-1} \Big(\frac{a}{b}\Big)^{2k - |\alpha |+1} 
 \big(\psi(1+k)+\psi(|\alpha |-k) \big);\\
 \gen\int_0^\infty  K_\alpha (br)^22r\dd
r =&\frac{(-1)^\alpha }{b^2} \Big( |\alpha 
     |\ln\big(\tfrac{b^2}{4}\big)+1
   +2|\alpha|\big(1-\psi(1+|\alpha|)\big)\Big).
\end{align*}

Similarly we  can be compute generalized bilinear
integrals of ${\bf
  S}_{\alpha,\ii\beta}$,   ${\bf Z}_{\alpha,\lambda}$. All these
formulas together with their derivations can be found in \cite{DeGaRu1}.

\section{Convergence of Gegenbauer functions and their integrals}

As $\beta,\lambda\to\infty$,  Gegenbauer functions converge to
Macdonald functions in the following sense:
\begin{align*}
\frac{\pi\e^{-\pi\beta}(\sin\theta)^{\alpha+\frac12}
}{2^\alpha\theta^{\alpha+\frac12}}
  {\bf S}_{\alpha,\pm\ii\beta}(-\cos\theta)  
 &=(\theta\beta)^{-\alpha} K_\alpha(\beta \theta)\big(1+O(\beta^{-1})\big); 
 \\
\frac{\sqrt\pi\Gamma(\tfrac12-\alpha+\lambda)(\sinh \theta)^{\alpha+\frac12}}{2^{\lambda+\frac12}  
\theta^{\alpha+\frac12}}
{\bf Z}_{\alpha,\lambda}(\cosh\theta)  
&=(\lambda\theta)^{-\alpha} K_\alpha(\lambda \theta)\big(1+O(\lambda^{-1})\big).
\end{align*}

The generalized  integrals of Gegenbauer functions converge to
the corresponding generalized integrals of Macdonald functions:
 \begin{align*}
 &\frac{\pi^2 {\e}^{-2\pi\beta} \beta^{2\alpha}}{2^{2\alpha}}
  \;\gen\int_{-2}^2{\bf S}_{\alpha,\ii\beta}(w)^2(1-w^2)^\alpha\dd 2 w
  =   \Big(1+\mathcal{O}\big(\tfrac{1}{\beta}\big)\Big)
    \;\gen\int_0^\infty  K_{\alpha}(\beta r)^2 2r \dd r;
\\
&\frac{\pi \Gamma\big(\tfrac12+\alpha+\lambda\big)^2
  }{2^{2\lambda+1}\lambda^{2\alpha}}
  \;\gen\int_2^\infty{\bf Z}_{\alpha,\lambda}(w)^2(w^2-1)^\alpha\dd 2 w 
 =  \Big(1+\mathcal{O}\big(\tfrac{1}{\lambda}\big)\Big)
    \;\gen\int_0^\infty  K_{\alpha}(\lambda r)^2 2r \dd r  .
 \end{align*}

 The convergence of these generalized integrals
 is straightforward in the non-anomalous case.  In the anomalous
case one has to choose the variables carefully, which we did: 
\begin{align*}
2r\dd r=\dd r^2,\quad 2(\cosh r -1)\simeq r^2,\quad
  2(1-\cos r)\simeq r^2.
\end{align*} 
 Note that the generalized integral is invariant with respect to the change of
variables $r \to r^2$, but not with respect to scaling. 
Proofs of these convergence statements can be found
in \cite{DeGaRu1}.

\section{Laplacian on the Euclidean space, the hyperbolic space and the sphere}

In the remaining part of our manuscript we describe an application of
generalized integrals to operator theory. These applications will
involve point interactions of the Laplacian.  We start by recalling 
some basic information about Green functions of Laplacians. The following 
formulas are well-known, confer for example \cite{CDT,DeGaRu2}. 

Consider the {\em Laplacian} $\Delta_d$ on the {\em Euclidean space} $\mathbb{R}^d$.
Let $G_d(z;x,x')$ be the {\em Euclidean Green function}, that is
the integral kernel of the resolvent
$    (-z-\Delta_d)^{-1}$. For $\Re\beta>0$, we have 
     \[
     G_d(-\beta^2;x,x')
     =\frac{1}{(2\pi)^{\frac{d}{2}}}\Big(\frac\beta{|x-x'|}\Big)^{\frac{d}{2}-1}K_{\frac{d}{2}-1}
         \big(\beta|x-x'|\big). \label{eq:flat_resolvent}
         \]

         The {\em hyperbolic space} is 
         \[\hh^d:=\{x\in\rr^{1,d}
         \mid[x|x]=1
         \}\]
  where       
         $[x|y]         =x^0y^{0}-x^1y^{1}-\dots-x^dy^{d}
         $ is the {\em Minkowskian  pseudoscalar product}.
         The hyperbolic distance between $x,x'\in\hh^d$ is given by 
         $\cosh(r)=[x|x']$.

                  Let $\Delta_d^\h$ denote the
         {\em Laplace-Beltrami operator} on $\hh^d$ and let
             $G_{d}^\h(z;x,x')$ be the {\em hyperbolic Green
               function},  that is, the integral kernel of
      $ \big(-z-\Delta_{d}^\h-\big(\tfrac{d-1}{2}\big)^2\big)^{-1}$. Then 
\[
     G_{d}^\h\Big(-\beta^2;x,x'\Big)
     =\frac{\sqrt\pi\Gamma(\frac{d-1}{2}+\beta)}{\sqrt2(2\pi)^{\frac{d}{2}}2^{\beta}
         }{\bf Z}_{\frac{d}{2}-1,\beta }                        \big([x|x']\big).\]

         The {\em unit sphere} is 
         \[\sS^d:=\{x\in\rr^{1+d}
         \mid(x|x)=1
                  \},\]where $(x|y)
                =x^0y^{0}+x^1y^{1}+\dots+x^dy^{d}$
                is the {\em Euclidean scalar product}.
         The spherical distance between $x,x'\in\sS^d$ is given by 
         $\cos(r)=(x|x')$. Let $\Delta_d^\s$ denote the
{\em         Laplace-Beltrami operator} on $\sS^d$. Let
             $G_{d}^\s(z;x,x')$ be the {\em spherical Green
               function}, that is, the integral kernel of  
             $ \big(-z-\Delta_{d}^\s+\big(\tfrac{d-1}{2}\big)^2\big)^{-1}$. Then
     \[
     G_{d}^\s(-\beta^2;x,x')
     =\frac{\Gamma(\frac{d}{2}-\frac12+\ii\beta )
       \Gamma(\frac{d}{2}-\frac12-\ii\beta )
     }{2^d\pi^{\frac{d}{2}}
         }{\bf S}_{\frac{d}{2}-1,\ii\beta }
         \big(-(x|x')\big).\]

       We remark that the integral kernels of spectral projections of
       the discussed operators  may be expressed explicitly in terms
       of special functions  of the same type as for resolvents, see e.g. \cite{DeGaRu2}.

       Let us also mention some basic properties of operators. 
 Let $H$ be a self-adjoint operator and let $G(-\rho):=(\rho+H)^{-1}$ be its resolvent. Then $G(-\rho)$ satisfies
\begin{align}\label{res1}
  (H+\rho)G(-\rho)&=\one,\\
  G(-\rho)^*&=G(-\bar \rho),\label{res2}\\
  \frac{\dd}{\dd \rho}G(-\rho)&=-G(-\rho)^2.\label{res3}
\end{align}
\eqref{res3} is called the {\em resolvent formula in the differential form.}

\section{Laplacian with point interactions}

Let us now try to define the Laplacian with a perturbation localized 
in a single point. For simplicity, we display details only for the 
case of the Euclidean space. We  look for an operator
$-\Delta_d^\gamma$, which is a self-adjoint extension of $-\Delta_d$ 
restricted to $C_\mathrm{c}^\infty(\rr^d\backslash\{0\})$.  The 
parameter $\gamma\in\rr\cup\{\infty\}$  will parametrize these 
self-adjoint extensions.

Actually, instead
of $-\Delta_d^\gamma$ it is more convenient to look for its resolvent
\begin{align*}
G_d^\gamma(-\rho )=(-\Delta_d^\gamma+\rho )^{-1}.
\end{align*}
By the conditions \eqref{res1}, \eqref{res2} and \eqref{res3}
 its integral kernel $G_d^\gamma(-\rho ,x,x')$ should satisfy
\begin{align*}(-\Delta_x + \rho )  G_d^\gamma(-\rho ,x,x')&=\delta(x-x'), \quad
                                                x\neq0,                                            \\
   G_d^\gamma(-\rho ,x,x')&=   G_d^\gamma(-\rho ,x',x),  \\
\partial_{\rho}   G_d^\gamma(-\rho,x,x')&=  -\int  G_d^\gamma(-\rho,x,y)
                                     G_d^\gamma(-\rho,y,x')\dd y. 
\end{align*}

These conditions are solved by a {\em Krein-type resolvent}
\begin{align*}
  G_d^\gamma(-\rho,x,x')&=   G_d(-\rho,x,x')+ 
  \frac{ G_d(-\rho,x,0)G_d(-\rho,0,x')}{\gamma+\Sigma_d(\rho)}, 
\end{align*}
where
\[
\partial_{\rho } \Sigma_d(\rho )  =\int_{\rr^d} G_d(-\rho ,0,y)^2\dd 
    y,\]
  and $\gamma$ is an arbitrary constant.
In dimensions $d=1,2,3$ the above integral is finite and we obtain
\[\Sigma_d(\beta^2)=\begin{cases}
    -\frac{1}{2\beta}&d=1;\\
    \frac{\ln(\beta^2)}{4 \pi}&d=2;\\
    \frac{\beta}{4\pi}&d=3.    
  \end{cases}\]
Thus we obtain formulas for the Green functions with a point
potential  in dimensions $d=1,2,3$ parametrized by a
real parameter $\gamma\in\rr$:
\[G_d^\gamma(-\beta^2;x,x')=\begin{cases}
    \frac{\e^{-\beta|x-x'|}}{2\beta}+\frac{\e^{-\beta|x|}\e^{-\beta|x'|}}{(2\beta)^2\big(\gamma-\frac{1}{2\beta}\big)},&d=1;\\[2ex]
    
    \frac{K_0(\beta |x-x'|)}{2\pi} 
+\frac{K_0(\beta |x|)K_0(\beta |x'|)}{(2\pi)^2(\gamma+\frac{\ln
    \beta ^2}{4\pi})} ,&d=2; 
    \\[2ex]
    
\frac{\e^{-\beta |x-x'|}}{4\pi|x-x'|}
+\frac{\e^{-\beta |x|}\e^{-\beta |x'|}}{(4\pi)^2 |x||x'| (\gamma+\frac{ \beta }{4\pi})} 
,&d=3.    \end{cases}\]
For dimensions $d=1,3$ we used the fact that for half-integer parameters 
the Macdonald function reduces to elementary functions.
Of course, the above construction is well known from the literature
\cite{AGHH,AK,BF}
and often used in the physics literature.

The operators $-\Delta_d^\gamma$, strictly speaking,  have  no analogs
for $d\ge4$.
However, the functions $G_d^\gamma(-\rho;x,x')$ can be generalized to $d\geq4$
using  {\em generalized integrals}: 
\[
  \partial_\rho\Sigma_d(\rho)= \frac{(\beta^2)^{\frac{d}{2}-1}\pi^{\frac{d}{2}}}{(2\pi)^d\Gamma(\frac{d}{2})}\gen\int_0^\infty
     K_{\frac{d}{2}-1}(\sqrt\rho r)^2 2r\dd r.\]

We obtain
\[
  \Sigma_d(\beta^2)=\begin{cases}
\frac{(-1)^{\frac{d+1}{2}}\beta^{d-2}}{(4\pi)^{\frac{d-1}{2}}2(\frac12)_{\frac{d-1}{2}}}&d\text{ odd};
    \\
\frac{ (-1)^{\frac{d}{2}+1}\beta^{d-2}}{(4
  \pi)^{\frac{d}{2}} \big( \frac{d}{2}-1 \big)!} \left( 2 -
2  \psi \big( \tfrac{d}{2} \big) + \ln \tfrac{\beta^2}{4}  \right)
&d\text{ even} .\end{cases}
\]
Thus for each dimension $d$ we obtain a family of Green functions
\begin{align}
\notag G_d^\gamma(-\beta^2;x,x')&
     =\frac{1}{(2\pi)^{\frac{d}{2}}}\Big(\frac\beta{|x-x'|}\Big)^{\frac{d}{2}-1}K_{\frac{d}{2}-1}
         \big(\beta|x-x'|\big)\\
                  +&
                    \frac{1}{(2\pi)^{d}}\Big(\frac{\beta^2}{|x||x'|}\Big)^{\frac{d}{2}-1}\frac{
                    K_{\frac{d}{2}-1}(\beta
                     |x|)K_{\frac{d}{2}-1}(\beta|x'|)}{\gamma+\Sigma_d(\beta^2)}.
\label{resol}\end{align}
describing  point interaction of strength controlled by  the 
parameter $\gamma$.
                 
A similar analysis can be performed for the hyperbolic and spherical 
Green functions in all dimensions \cite{DeGaRu2}.

                 One can ask what is the meaning of
 $G_d^\gamma(-\beta^2;x,x')$ in dimensions $d\geq4$, for wich it is
 not the kernel of a bounded operator  (and in particular, not a kernel
 of a resolvent of a self-adjoint operator). Our expectation is  as follows.
Suppose that $V$ is a {\em potential} on $\rr^d$, $\hh^d$ or $\sS^d$, possibly strong but  with a small
support. Consider the
the {\em Schr\"odinger operator}  $-\Delta_d+V$.
Let \[G^V(-\beta^2)=(\beta^2-\Delta_d+V)^{-1}\]
be its resolvent with the integral kernel
$G^V(-\beta^2;x,x')$.  Then far from the support of $V$
we can approximate $G^V(-\beta^2;x,x')$ by $G^\gamma(-\beta^2;x,x')$
as in \eqref{resol},
possibly adding to $\Sigma_d(\beta^2)$
a  polynomial in the energy
of degree $<\frac{d-1}{2}$ if $d$ is odd and  $<\frac{d-2}{2}$ if $d$
is even. Thus possible infrared behaviors of $G^V(-\beta^2;x,x')$ are
controlled by a finite number of parameters (coefficients of the above
mentioned polynomials).

The choice given by the zero polynomial that is defined by the above
generalized integrals can be viewed as a ``standard reference point''. 
The generalized integral computed in different coordinates, which are 
related to $r^2$ by a well-behaved coordinate transformation in the sense of 
\eqref{eq:change_of_var}, will yield a different choice of polynomial. 

The observation that a well-behaved change of variables will, by \eqref{eq:change_of_var}, only change this polynomial is particularly important 
in the curved cases, where we chose $2(\cosh r -1)$, resp. $2(1-\cos r)$ as 
integration variables instead of $r^2$.

This is  analogous to the renormalization of quantum quantum field theory using 
dimensional regularization, see e.g. \cite{BG96}.

The above situation resembles  the idea often expressed
in the context of {\em quantum field theory} and of
the theory of {\em critical phenomena},  attributed to Keneth Wilson: for large distances
correlation functions have a {\em universal behavior} independent of the
details of the interaction, described by few parameters.

\ack
The work of J.D. and G.G. was supported by National Science Center
(Poland) under the Grant UMO-2019/35/B/ST1/01651.

\end{document}